\documentclass[twocolumn, tighten, times]{aastex63}
\usepackage{amsmath,amstext}
\usepackage[T1]{fontenc}
\usepackage[figure,figure*]{hypcap}
\usepackage{multirow}
\usepackage[T1]{fontenc} 
\usepackage[utf8]{inputenc} 
\usepackage{float}
\newcommand{\ba}{\begin{eqnarray}}
\newcommand{\ea}{\end{eqnarray}}
\usepackage{lineno}
\usepackage{soul}

\received{XXX}
\accepted{YYY}

\shorttitle{Planetesimal Transport}
\shortauthors{M. Best, A. Sefilian, \&  C. Petrovich}

\begin{document}

\title{The influence of cold Jupiters in the formation of close-in planets. I. planetesimal transport}

\correspondingauthor{Marcy Best}
\email{sabest@uc.cl}

\author[0000-0001-8361-9463]{Marcy Best}
\affiliation{Instituto de Astrofísica, Pontificia Universidad Católica de Chile, Av. Vicuña Mackenna 4860, 782-0436 Macul, Santiago, Chile}
\affiliation{Millennium Institute of Astrophysics MAS, Nuncio Monsenor Sotero Sanz 100, Of. 104, Providencia, Santiago, Chile}

\author[0000-0003-4623-1165]{Antranik A.  Sefilian}
\altaffiliation{Alexander von Humboldt Postdoctoral Fellow.}
\affiliation{Astrophysikalisches Institut und Universit{\"a}tssternwarte, Friedrich-Schiller-Universit\"at Jena, Schillerg\"a{\ss}chen~2--3, D-07745 Jena, Germany}

\author[0000-0003-0412-9314]{Cristobal Petrovich}
\affiliation{Instituto de Astrofísica, Pontificia Universidad Católica de Chile, Av. Vicuña Mackenna 4860, 782-0436 Macul, Santiago, Chile}
\affiliation{Millennium Institute for Astrophysics, Chile}

\begin{abstract}
\noindent
The formation of a cold Jupiter (CJ) is expected to quench the influx of pebbles and the migration of cores interior to its orbit, thus limiting the efficiency of rocky planet formation either by pebble accretion and/or orbital migration. Observations, however, show that the presence of outer CJs ($>1$ au and $\gtrsim 0.3M_{\rm Jup}$) correlates with the presence of inner Super Earths (at $<1$ au). This observation may simply be a result of an enhanced initial reservoir of solids in the nebula required to form a CJ or a yet-to-be-determined mechanism assisted by the presence of the CJ. In this work, we focus on the latter alternative and study the orbital transport of planetesimals interior to a slightly eccentric ($\sim0.05$) CJ subject to the gravity and drag from a viscously-evolving gaseous disk. We find that a secular resonance sweeping inwards through the disk gradually transports rings of planetesimals when their drag-assisted orbital decay is faster than the speed of the resonance scanning. This snowplow-like process leads to large concentration (boosted by a factor of $\sim 10-100$) of size-segregated planetesimal rings with aligned apsidal lines, making their expected collisions less destructive due to their reduced velocity dispersion. This process is efficient for a wide range of $\alpha-$disk models (and thus disk lifetimes) and Jovian masses, peaking for $\sim 1-5 M_{\rm Jup}$, typical of observed CJs in radial velocity surveys. Overall, our work highlights the major role that the disk’s gravity may have on the orbital redistribution of planetesimals, depicting a novel avenue by which CJs may enhance the formation of inner planetary systems, including super-Earths and perhaps even warm and hot Jupiters.


\vspace{0.75cm}

\end{abstract}

\section{Introduction}

\begin{figure*}[]\center \label{fig:schematic}
\includegraphics[width=9cm]{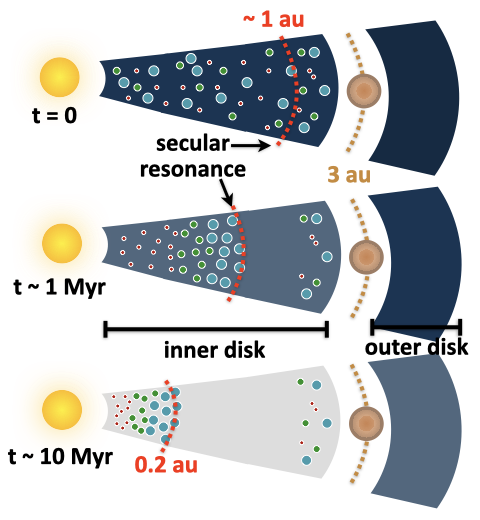}
\caption{Schematic cartoon describing the basic setup of our simulations.  A giant planet carves a gap in a depleting gaseous disk with a planetesimal disk interior to its orbit. 
As the disk disperses, a secular resonance that excites the planetesimals' eccentricities moves inwards carrying them along due to gas drag,  accumulating  material in the form of size-segregated  planetesimal rings. The final surface density of planetesimals inside $\sim 0.2$ au  is then largely enhanced.}	
\end{figure*}

\subsection{Motivation}

Recent exoplanet observations show that the presence of cold giant planets may increase the probability of finding an inner super-Earth in the same system
\citep{correlation1,correlation2}. These works estimate that the probability of having at least one super-Earth given a cold giant in the exoplanetary system is $\sim 50-100\%$ (within 1$\sigma$; more recently \citealt{correlation_2021} also found such a correlation, albeit less significant). Nevertheless, this trend is a subject of current debate,  with other works claiming the contrary, i.e., an anti-correlation \citep[e.g.][]{bonomo}. Thus, understanding the role that cold Jovians have at promoting (or inhibiting) the formation of inner super-Earths is a key step forward in building a complete picture of planet formation. 

According to some formation models the presence of a cold Jupiter is seen as an obstacle to super-Earth formation. These models generally rely on the transport of rocky material (e.g., pebbles or cores) from the outer regions of the disk to the inner regions to form super-Earths. This is why, usually, an anti-correlation is predicted (e.g. \citealt{survey}) between super-Earths and Cold Jupiters.

Indeed, in the pebble accretion model, planets form from planetesimals that accrete pebbles until they become sufficiently large (\citealt{pebble1}; \citealt{pebble2}). However, if a giant planet is formed early on in the outer regions of the protoplanetary disk, it would quench the flux of pebbles to the inner system. This is because the typical core of the Jovian reaches the pebble isolation mass \citep{Lambrechts_2014}. This blockage should halt the accretion of pebbles by the inner planetesimals, limiting the formation of other planets interior to the initial gas giant as the remaining pebbles would just drift towards the star too quickly. Also, the same pressure bump aiding the formation of  the gas giant would quench the pebble flux \citep{izidoro2021}.

Instead of pebbles, the influx of solids may be dominated by cores undergoing type-I migration that park into super-Earth orbits. However, the formation of a giant planet would also prevent their migration, giving rise to an anti-correlation between the two populations \citep{izidoro2015}.

Finally, another possible explanation for the observed correlation is that the presence of the cold Jupiter has no significant effect on the formation of a super-Earth. Instead, the potential correlation simply emerges as a consequence of birth disk conditions: disks with more solid material inside $\sim 10$ au may promote the formation of both populations (see e.g. the discussion in \citealt{Zhu2018}). This can happen either by having a more massive disk or having a higher dust-to-gas ratio  (i.e. more solid material to form planets). Although there is still debate about the exact distribution of material in the protoplanetary disk \citep{mmen}, the distribution of solids in multiple planet systems seems to be significantly different from what models of gaseous disks can explain. This points towards a rearrangement of the solid material \citep{mmen_controversy}.

In this work, we explore yet another possibility where the same giant planet blocking the pebbles or inwardly migrating planetary cores could transport copious amounts of planetesimals from $\gtrsim1$ au to the inner regions. This would imply that terrestrial planets in such a scenario grow by accreting planetesimals, possibly assisted by the higher density of solids required to form the Jovian.

\subsection{Sweeping secular resonances}

The mechanism that we invoke to transport planetesimals inwards is based on the sweeping of secular resonances to drive eccentricity excitation and gas drag to damp them, causing orbital decay. The sweeping arises from the commensurability between the apsidal precession rates of the Jovian planet and that of the planetesimals (\citealt{hep80}; \citealt{war81}), a process that has been studied in various contexts, including  the dynamics of asteroids in the Solar System \citep{lemaitre1991,nagasawa99,zheng17a,zheng17b,gong19} and the excitation of planetary eccentricities and/or inclinations \citep{hahn99,nagasawa03,petro2019,Toliou2019}. 

This process can be understood as follows (see also Figure \ref{fig:schematic} for an illustration). The gap-opening Jovian planet will undergo apsidal precession with a positive frequency due to the disk's gravity \citep{war81}. On the other hand, non-gap-opening planets/planetesimals will precess (generally) at a negative rate due to the disk \citep{war81} or positively due an external perturber like in the case of planetesimals affected by the Jovian(s). The depletion of the disk can lead to instances when the frequencies match at a given orbital radius, driving the excitation of the planetesimals' eccentricities when a slightly eccentric giant planet is considered (see Figure \ref{fig:fig1} for an illustration). These eccentricities are damped by the drag from the gas, pushing the planetesimals inwards, plowing them through the inner system as the resonance moves inward as the disk disperses. This can effectively transport a large percentage of solid material available in the region.

A major goal of our work is to understand these dynamics for realistic disks by accounting for gaps, inner cavities, and different disk viscosities; expanding previous treatments which typically use uniform power-law disks. 

\subsection{Structure}

Our work is organized as follows. In Section \ref{sec:methods} we describe the general setup of our model, and present equations governing the dynamics of planetesimals.  In Section \ref{sec:transport} we then present results showing the transport of material due to our proposed mechanism, while the impact the alignment of planetesimals has on their velocity dispersion is presented in in Section \ref{sec:velocity}. Finally we discuss some of the implications our findings may have for planet formation in Section \ref{sec:discussion} before giving our conclusions in Section \ref{sec:conclusions}.

 \section{Methods} 
\label{sec:methods}

In this section, we describe a semi-analytic model to analyze the long-term dynamical evolution of planetesimals that are embedded within a viscously evolving protoplanetary disk and perturbed by a single planet. 

The planet, the disk, and the planetesimals within it are assumed to be coplanar orbiting a central star of mass $M_c = 1M_{\Sun}$. We characterize the orbit of a given planetesimal by its semimajor axis $a_{\rm p}$, eccentricity $e_{\rm p}$, and longitude of pericenter $\varpi_{\rm p}$. Similarly, orbital elements subscripted with J refer to the accompanying giant planet.

In order to follow the evolution of the planetesimals we need to model their interaction with the giant planet. Additionally, we need to include the effects of the gaseous disk component. This component produces gas drag onto the planetesimals, and drives apsidal precession of the planetesimals and the giant. The following equations describe all the interactions considered:

\begin{eqnarray}
\frac{de_{\rm p}}{dt} &=& \left( \frac{de_{\rm p}}{dt} \right)_{\rm drag} + \left( \frac{de_{\rm p}}{dt} \right)_{\rm CJ} , \label{eq:1}
\\
\frac{d\Delta \varpi}{dt} &=& \left( \frac{d\varpi_{\rm p}}{dt} \right)_{\rm disk} + \left( \frac{d\varpi_{\rm p}}{dt} \right)_{\rm CJ} - \left( \frac{d\varpi_{\rm J}}{dt} \right)_{\rm disk}  , \label{eq:2}
\\
\frac{da_{\rm p}}{dt} &=& \left( \frac{da_{\rm p}}{dt} \right)_{\rm drag} , \label{eq:3} 
\end{eqnarray}
where $\Delta \varpi = \varpi_{\rm p} - \varpi_{\rm J}$.

In subsequent subsections, we describe in detail each of these interactions, as well as the evolution of the gaseous component of the disk and the initial distribution of planetesimals.

\subsection{Disk evolution} \label{sec:disk_evolution}

We consider a gaseous protoplanetary disk that evolves as an $\alpha$ accretion disk, without considering the influence of any other physical effects (e.g. due the giant planet, photoevaporation winds, etc). In our most general case, two different viscosities are considered (to the inner and outer side of the giant planet), which is further described in Section \ref{sec:viscosity}. We compute the time evolution of the gas surface density $\Sigma_{\rm gas}$ using the classical 1D diffusion equation \citep{Pringle1981}:

\begin{eqnarray}\label{eq:viscosity}
    \frac{\partial\Sigma_{\rm gas}}{\partial t} &=& \frac{1}{r} \frac{\partial}{\partial r} \left[ 3r^{1/2} \frac{\partial}{\partial r}(\nu \Sigma_{\rm gas} r^{1/2}) \right] ,
 \\
    \nu &=& \alpha c_s H   , 
\end{eqnarray}
and account for a gap profile produced by the giant planet a posteriori; see Section \ref{sec:gap} for details.
Here, $\alpha$ is the viscosity parameter, $c_s$ is the sound speed, and H is the scale height, which in our case of a locally isothermal disk, correspond to $c_s \propto r^{-1/4}$ and $H \propto r^{5/4}$ (with $H \approx 0.04$ au at $1$ au).

\subsubsection{Initial profile}

We consider a gas surface density which initially has a power-law profile with a cutoff at some outer radius:

\begin{equation} \label{eq:initial_profile}
    \Sigma_{\rm gas}(r) =  \Sigma_{\rm gas,0} \left(\frac{r}{1 \rm au}\right) ^ {-\gamma} \exp\left[-(r/r_{\rm cut})^{2-\gamma}\right],
\end{equation}
where $\Sigma_{\rm gas,0}$ is chosen so that the total mass of gas is 1\% of the stellar mass. Motivated by observations, we adopt a $\gamma$ index of 1 here and $r_{\rm cut}$ to be 25 au, although our results depend only slightly on this parameter. Here, we note that variations in these parameters do not affect the evolution of planetesimals significantly as the shape of the profile reaches a steady state irrespective of the initial value of $\gamma$ (see Appendix \ref{app:evolution}). Additionally, the planetesimals/planet's orbital precession depends more on the total mass of the outer disk and not so much on its outer edge (see Appendix \ref{app:resonance}).

\subsubsection{Viscosity} \label{sec:viscosity}

The evolution of the disk's surface density depends on the gas viscosity; see e.g. Eq. (\ref{eq:viscosity}). This directly impacts the rate of depletion of a region; namely, the higher the viscosity, the quicker the region depletes its gas. At an iceline or the boundary of a dead zone, it is generally thought that the gas viscosity changes\footnote{In Fig. \ref{fig:A1} we see a surface density transition at this location.}, serving as a dust trap where a giant planet can form early on \citep{guilera2020}. Following this line of reasoning, we use the same position for both the viscosity transition and the position of the giant planet (namely, $a_{\rm J} = 3$ au, unless stated otherwise).
In this work we will refer to these viscosities as $\alpha_{\rm in}$ and $\alpha_{\rm out}$ (inside and outside the boundary, respectively). This gives us the freedom of separating the depletion of the inner and outer disks which, as we shall see, has important consequences for the evolution of the secular resonance location (Appendix \ref{app:resonance}).

\subsubsection{Density gap opened by the Jovian} \label{sec:gap}

As the planet accretes the surrounding gas, it will open a gap in the gas disk if it is sufficiently massive (see the red region of panels B1 to B3 in Fig. \ref{fig:2}). We need to keep track of this effect, as the drag force and precession rate of the planetesimals depend on the gas surface density. In principle, this can be done self-consistently by adding a torque at the position of the giant planet to the viscous evolution equation (Eq. (\ref{eq:viscosity})). For simplicity, however, we model the gap using the profile and code provided in \cite{Duffell_2020} for each time step independently. For reference, an approximate expression for the gap width $\Delta_{\rm gap}$ is provided by Equation (4) of \cite{Kanagawa2016}:
\begin{equation} \label{gap}
    \frac{\Delta_{\rm gap}}{a_{\rm J}} = 0.41 \left( \frac{M_{\rm J}}{M_c} \right) ^ {1/2} h ^ {-3/4} \alpha^{-1/4},
\end{equation}
where $M_{\rm J}$ is the mass of the giant, $h$ is the disk's aspect ratio at the position of the Jovian, and the viscosity of the inner disk ($\alpha_{\rm in}$) is considered for the gap. We use this simplified expression for our model in Appendix \ref{app:resonance}.

\subsection{Drag force}

The planetesimals embedded in the disk feel a drag force when the gas in their vicinity moves at a non-Keplerian speed. We can quantify this difference in velocity with the factor $\eta(r)$.
From hydrostatic equilibrium, a disk with volumetric gas density
$\rho\propto r^{-\Gamma}$ and temperature $T\propto r^{-\beta}$ has an azimuthal velocity given by \citep[e.g.][]{adachi76}:
\begin{equation}
v_{\rm gas}(r)=v_{\rm K}(r)\left[1-2\eta(r)\right]^{1/2}\simeq v_{\rm K}(r)\left[1-\eta(r)\right],
\end{equation}
with
\begin{equation}
\eta(r)=\frac{\pi(\Gamma+\beta)}{16}\frac{c_m^2}{v_K^2} = \frac{(\Gamma+\beta)}{2}h^2,
\end{equation}
where $v_K$ is the Keplerian velocity, and we have used the fact that $c_s/v_K$ is equal to the aspect ratio of the disk $h$, and the mean thermal speed is $ c_m^2 =\frac{8}{\pi} c^2_s$.

In our models, we use $\beta = 1/2$ and we can obtain $\Gamma$(r) locally\footnote{Computing $\Gamma$(r) locally is important because the surface density quickly deviates from a power law due to the disk evolution, especially near the edges of the disk, the gap of the planet, and the viscosity transition.} by calculating:

\begin{equation}
    \Gamma(r) = -\frac{r}{\rho} \frac{d\rho}{dr}.
\end{equation}
By averaging the drag force over one orbit, \citet{adachi76} find the following approximate expressions for the migration of the planetesimals:
\begin{align} \label{eq:migration}
\left( \frac{da_{\rm p}}{dt} \right)_{\rm drag} &= -\dfrac{2a_{\rm p}}{\tau_0} \sqrt{\dfrac{5}{8}e_{\rm p}^2+\eta^2} \left[ \eta +\left( \dfrac{\Gamma}{4} + \dfrac{5}{16} \right) e_{\rm p}^2 \right],
\\
\left( \frac{de_{\rm p}}{dt} \right)_{\rm drag} &= -\dfrac{e_{\rm p}}{\tau_0} \sqrt{\dfrac{5}{8}e_{\rm p}^2+\eta^2}, \label{eq:ecc_evolution}
\end{align}
where 
\begin{align}
\tau_0 &= \left( \dfrac{\pi C_{\rm D}}{2m}s^2\rho v_K \right)^{-1}. \nonumber
\end{align}
where $s$ is the radius of the planetesimal, $C_D$ the drag coefficient, and $\rho$ the density of planetesimals which we take to be 1 g/cm$^3$. Note that we are not considering the inclinations of the planetesimals in our work, as these should be suppressed due to the Jovian's inclination being close to zero.

\subsection{Apsidal precession due to the disk gravity}

Next, we consider the gravitational effects of the protoplanetary disk on both the planetesimals embedded within and the planet. Given that the disk is taken to be axisymmetric,  it simply causes the longitudes of periapsis of the planetary and planetesimal orbits to precess.

Since these objects are surrounded by the disk, the gas can be infinitely close, leading to a numerical (not physical) problem which causes a divergence in the gravitational force \citep[see][for a detailed discussion]{SR2019}. To overcome this problem, we soften the gravitational potential using the softening prescription of \citet{Hahn_2003} so that \citep{SR2019}:
\begin{equation}
    \dfrac{1}{|\textbf{r}-\textbf{r'}|} = \left[ r^2+r'^2 - 2rr'\cos(\theta) + h^2(r^2+r'^2) \right]^{-1/2},
\end{equation}
Accordingly, the usual Laplace coefficients,
\begin{eqnarray}
    b_{\rm s}^{\rm (m)}(\alpha) &=& \frac{2}{\pi} \int_{\rm 0}^{\pi} \frac{\cos(m\theta) d\theta}{\big(1+\alpha^2 - 2\alpha\cos\theta \big)^s}  , 
\end{eqnarray}
are modified which now read as follows:
\begin{eqnarray}
    \mathcal{B}_{\rm s}^{\rm (m)}(\alpha) &=& \frac{2}{\pi} \int_{\rm 0}^{\pi} \frac{\cos(m\theta) d\theta}{\big(1+\alpha^2 - 2\alpha\cos\theta + (1+\alpha^2)h^2 \big)^s}  .\nonumber \\
\end{eqnarray}
Here, we note that the aspect ratio is taken to be a distance-independent constant in \citet{Hahn_2003}.
In our model, however, we assume that $h$ varies with semimajor axis. As a consequence, we use its value at the position of the object. This approximation should be valid since the smoothening only affects the potential in the neighbourhood of the object.

Then, we can calculate the precession due to an axisymmetric disk using the following equation from \cite{SR2019}:
\begin{align}
    \nonumber
    \left( \frac{d\varpi_i}{dt} \right)_{\rm disk} = \frac{2G}{n_i a_i^3} & \bigg[  \int_{\rm r_{\rm in}}^{\rm a_i} \mu(r) \phi\left(\frac{r}{a_{\rm i}}\right)dr
\\
    &+ \int_{\rm a_i}^{\rm r_{\rm out box}} \frac{a_{\rm i}}{r} \mu(r) \phi\left(\frac{a_{\rm i}}{r}\right)dr \bigg] , \label{eq:dwi_dt_disk}
\end{align}
where
\begin{align}
    \phi(x) = \frac{1}{8} x \left[ \mathcal{B}_{\rm 3/2}^{(1)}(x) - 3 x h^2(2+h^2)\mathcal{B}_{\rm 5/2}^{(0)}(x) \right],
\end{align}
where i = \{p, J\},  $n_i = \sqrt{G (M_c+m_i) / a_i^3}$ its mean motion, $r_{\rm out, box}$ the outer limit of the simulation box (1000 au) and $\mu(r)$ is the gas mass density per unit semimajor axes, $\mu(r) = 2 \pi r \Sigma_{\rm gas}$. Equation (\ref{eq:dwi_dt_disk}) generally gives a negative value for the planetesimals (subscript p) because they are embedded in the gas, but a positive value for the giant planet (subscript J) because of the gap the latter opens \citep[see e.g.][]{ward1981, SR2019}.

It is worth noting that the singularity discussed above does not arise when considering objects orbiting outside the disk, i.e., where $\Sigma_{\rm gas}(r) = 0$, such as the planet in our model.
We also  point out that there are analytic frameworks that allow for the computation of the disk-induced precession without introducing any softening parameter (see e.g. \citet{davydenkova2018}). Despite this, however, we opted to adopt the softening prescription of \citet{Hahn_2003}, as the resulting softening parameter $h$ is the same as the disk's  aspect ratio. This has the added effect of accounting for the gaseous disk not being razor-thin, which is a more realistic representation.

\subsection{Planetesimal-Jupiter interaction}

We now consider the orbital evolution of the planetesimals due to the Jovian. Making use of the classical disturbing function $\mathcal{R}$ expanded to second order in eccentricities (e.g. Eq. (7.8) in \cite{MD2000}), one can write:

\begin{align}
    \left( \frac{de_{\rm p}}{dt} \right)_{\rm CJ} &= \dfrac{1}{n_{\rm p}a_{\rm p}^2e_{\rm p}}\dfrac{\partial \mathcal{R}}{\partial \varpi_{\rm p}} = A_{\rm pJ} e_{\rm J} \sin(\varpi_{\rm p} - \varpi_{\rm J}), \label{eq:ecc_evolution_2}
    \\
    \left( \frac{d\Delta\varpi}{dt} \right)_{\rm CJ} &= \dfrac{1}{n_{\rm p}a_{\rm p}^2e_{\rm p}}\dfrac{\partial \mathcal{R}}{\partial e_{\rm p}} = A_{\rm pp} + A_{\rm pJ}\dfrac{e_{\rm J}}{e_{\rm p}}\cos(\varpi_{\rm p} - \varpi_{\rm J}),\label{eq:ecc_evolution_3}
\end{align}
where  $n_{\rm p}$ is the planetesimal's mean motion. In Equations (\ref{eq:ecc_evolution_2}) and (\ref{eq:ecc_evolution_3}), the coefficients $A_{\rm pp}$ and $A_{\rm pJ}$ are given by:
\begin{eqnarray}
A_{\rm pp} &=& \dfrac{n_{\rm p}}{4}\dfrac{M_{\rm J}}{M_{\rm c} + m_{\rm p}} \left(\frac{a_{\rm p}}{a_{\rm J}}\right)^2 b_{\rm 3/2}^{(1)}(a_{\rm p}/a_{\rm J}), 
    \\
A_{\rm pJ} &=& -\dfrac{n_{\rm p}}{4}\dfrac{M_{\rm J}}{M_{\rm c} + m_{\rm p}} \left(\frac{a_{\rm p}}{a_{\rm J}}\right)^2 b_{\rm 3/2}^{(2)}(a_{\rm p}/a_{\rm J}),
\end{eqnarray}
see e.g. \citet{MD2000}. It is worth noting, that in the case of a circular orbit, the Jovian would not be able to secularly excite any eccentricity onto the planetesimals. Thus, the eccentricity of the giant is an important parameter in our simulations, and so, we have tested different values in Section \ref{sec:others}.

\subsection{Secular Resonance}

\begin{figure}[]\label{fig:fig1}
\hspace{0.085cm}
\includegraphics[width=10cm]{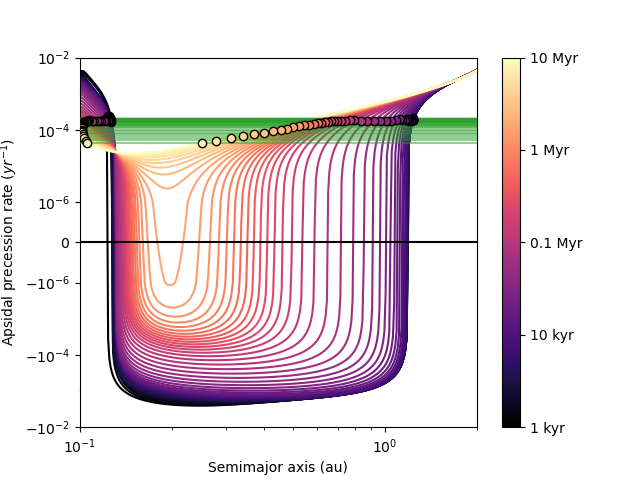}
\caption{Precession rate of the planetesimals due to both the disk and planet gravity as a function of semimajor axis. The different lines are for different times as described by the colorbar. The green lines represent the precession rate of the giant planet which decreases in time as the disk depletes. For reference, the resonance position as a function of time is marked with a circle symbol corresponding to the same colorbar. The resonance position moves from approximately 1.5au to 0.2au at the end of the 10 Myrs. A second resonance can also be seen at the inner edge, but this one spans a narrow range from 0.1 au to 0.15 au.  The resonance position as a function of time can be found in Fig. \ref{fig:7}.}
\end{figure}

Now that all the sources for the apsidal precession of the Jovian and planetesimals are defined, we proceed to describe when a secular resonance takes place. Loosely speaking, this kind of resonance occurs when we have a matching of the apsidal precession rates (in our case between the planet and some planetesimals). This would mean that Eq. (\ref{eq:2}) is close to zero in the absence of the gas drag which aligns (or anti-aligns) the orbits. A planetesimal in this secular resonance will increase its eccentricity up to a maximum value (see Eq. (\ref{eq:peak_ecc}) for an estimate). Once the planetesimal acquires enough eccentricity, its migration will be significant (Eq. (\ref{eq:migration})). For illustrative purposes, in Fig. \ref{fig:fig1} we plot the precession rate of the planetesimals in the fiducial case evolving in time as a function of semimajor axis. Also, overlapping, is the precession rate of the Jovian (in green) which decreases in time as the disk depletes. We show the matching of the frequencies as a circle at each time. As one can see, the resonance starts close to $1.5$ au and moves inwards towards the inner edge at $0.1$ au. In the next section (Section \ref{sec:transport}), we shall demonstrate how this sweeping of the resonance could transport material.

\subsection{Planetesimal distribution}

Finally, with the basic elements of the model in place, we describe how the planetesimals themselves are distributed throughout the disk. To allocate the number of planetesimals at the beginning of the simulation, we consider 200 logarithmically spaced rings spanning from 0.1 au to 1.5 au. Each ring contains planetesimals such that their combined mass is 1\% of the corresponding gas mass for the ring. In turn, the size distribution for planetesimals follows a power law:

\begin{equation} \label{eq:planetesimals}
    \frac{dN}{dM} \propto M^{\rm -p},
\end{equation}
where N is the number of planetesimals in each of the 50 logarithmically spaced mass bins M (from 1 km to 100 km with a density of 1 g/cm$^3$). Looking at Eq. (\ref{eq:planetesimals}), it can be easily seen that with $p = 2$, all sizes of planetesimals contribute to the mass budget with the same total mass. Values below 2 are top heavy (meaning, the bigger planetesimals contribute more to the mass budget), while values above 2 are weighted towards the smaller planetesimals. In what follows we use $p = 2$ as to not prioritize any size of planetesimals. The exact distribution of planetesimals will not significantly affect our results as we will discuss the evolution of each size of planetesimals in what follows.

Our choice for the distribution in semimajor axis of the planetesimals is 0.1 au to 1.5 au. The lower limit is simply the inner edge of the disk and the upper limit is introduced because planetesimals close to the planet would be evactuated by the mean motion resonances (MMRs). We discuss this further in Section \ref{sec:mmr}.

\begin{figure*}[]\center \label{fig:2}
\includegraphics[width=18cm]{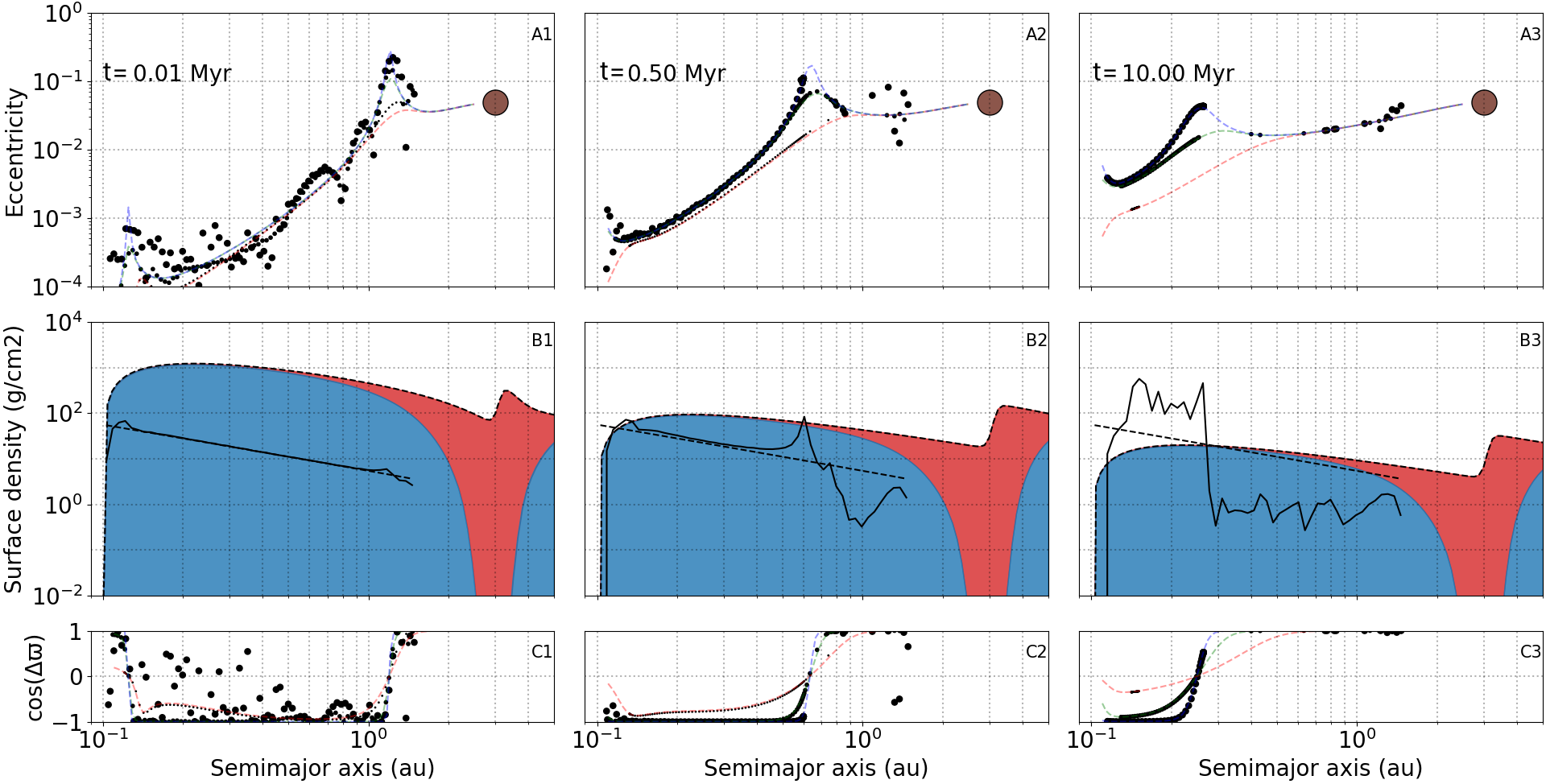}
\caption{Evolution of the planetesimals and the gaseous disk. Row A shows the eccentricity of planetesimals as a function of semimajor axis for a sub-sample of sizes ($s = 1$ km, $10$ km, and $100$ km respectively, shown in black circles of increasing size). For reference, the corresponding forced eccentricity solutions obtained by solving Equations (\ref{eq:1}) and (\ref{eq:2})) are also shown in red, green, and blue dashed lines, respectively. 
Row B shows the gas surface density in blue with the gap carved by the planet in red. The entirety of the gap is not shown here for aesthetic reasons, as the surface density at $a_p$ goes down to $\Sigma_{\rm gas} \sim 10^{-5} {\rm g.cm^{-2}}$ at $t = 0$. The solid black line represents the surface density of planetesimals, and for reference the initial profile is shown with a dashed black line. 
Row C shows the cosine of the relative apsidal angle of planetesimals with respect to the precessing giant planet. 
For reference, the forced apsidal angles are also shown using the same color scheme as Row A. Each column corresponds to a different time: from left to right, $t = 0.01$, $0.5$, and $10$ Myrs. See Section \ref{sec:fiducial} for further details. An animated version of this figure is available in the electronic edition of the journal and in Figshare: \href{http://doi.org/10.6084/m9.figshare.24132339}{10.6084/m9.figshare.24132339}. The animation runs from t = 0 to t = 10 Myrs with a duration of 13 s.} 	
\end{figure*}

\section{Results: Transport}
\label{sec:transport}

\subsection{Fiducial case} \label{sec:fiducial}

\begin{table}[t!]
\caption{Initial conditions of the disk and planet for our simulations.
\label{table:models}}

\begin{tabular}{ccc}
\tableline
\tableline

Parameter & Fiducial value & Other values \\

\tableline

$M_{\rm J} [M_{\rm Jup}]$ & 3 & [0.1 -- 7] \\

$a_{\rm J} [au]$ & 3 & 5 \\

$e_{\rm J}$ & 0.05 & 0.01 \\

$\alpha_{\rm in}$ & $10^{-3}$ & [$10^{-4}$ -- $10^{-2}$], [$3 \times 10^{-3}$] \\

$\alpha_{\rm out}$ & $10^{-4}$ & [$10^{-5}$ -- $10^{-3}$],[$3 \times 10^{-3}$] \\

\tableline
\end{tabular}

\tablecomments{In each simulation, unless stated otherwise, (i) the surface density slope is set to $\gamma=1$ (Eq. (\ref{eq:initial_profile})); (ii) the initial gas mass is set to 1\% of solar mass; (iii) the planetesimal size distribution slope is set to $p = 2$ (Eq. (\ref{eq:planetesimals})); (iv) the total solid mass in the inner disk is set to 1\% of the gas mass; and (v) the simulation time is 10 Myr.}
\end{table}

We simulate the evolution of planetesimals ranging from 1 km to 100 km, keeping track of their semi major axes, eccentricities, and apsidal angles (with respect to the Jovian) for $10$ Myrs; see Equations (\ref{eq:1})--(\ref{eq:3}). All parameters considered for the fiducial case are listed in Table \ref{table:models} and in text form here. In the fiducial case, we invoke a giant planet of 3 Jupiter masses (a typical cold giant planet) at 3 au (the same position as the snowline), and $e_{\rm J}=0.05$ (which is reasonable as the planet's eccentricity can grow up to the order of the local aspect ratio \citep{papaloizou,Duffell_2015}). The inner and outer viscosity parameters for the disk are $\alpha_{\rm in}=10^{-3}$ and $\alpha_{\rm out}=10^{-4}$. We start the simulation with a gas disk mass of 1\% of a solar mass and a solid surface density of 1\% of the gas disk mass. As planetesimals do not interact among themselves, the amount of available solid material can be easily scaled up or down later and we shall express our results in fraction of transported mass.

The results of our fiducial simulation are summarized in Figure \ref{fig:2}, where we plot the radial profiles of the planetesimal eccentricities (row A), surface densities of the gas and planetesimals (row B), and planetesimal apsidal angles (relative to the precessing planet; row C). Results are shown at three different times: from left to right, $ t = 0.01$, $t = 0.5$, and $t = 10$ Myr, respectively. For clarity purposes, only a sub-sample of three planetesimal sizes are shown: namely, 1, 10, and 100 km. Starting from panel A1 we see the planetesimals  oscillating around the forced eccentricity profiles (dashed lines). These forced eccentricities are calculated by simultaneously solving Eq. (\ref{eq:1}) and (\ref{eq:2}) looking for a steady state (which is different for each planetesimal size). The different terms for these equations are described throughout Section \ref{sec:methods}. We consider each time step to be independent, using the previously calculated gas surface density as described in Section \ref{sec:disk_evolution}. As the resonance location moves towards the star, it carries the planetesimals with it (panels A1 to A3), thus sweeping up the solid material. Whether the planetesimals move with the resonance or not is explored in Section \ref{sec:condition}. Note the corresponding increase in solid surface density (black line) in panels B1 to B3. We will quantify this increase in surface density in Section \ref{sec:amount}.

Some things to further note. First, the largest planetesimals of 100 km depend on the resonance position to migrate, while the smaller ones of 1 km migrate almost irrespective of the resonance position, as the planet's perturbation already excites eccentricities that are large enough to drive their migration (see Section \ref{sec:condition}).

Second, in panels C1 to C3, we see planetesimals close to the planet tend to become aligned with the orbit of the planet (i.e., $\Delta\varpi = 0$), while planetesimals further away tend to become anti aligned (i.e., $\Delta\varpi = \pi$). This means that there is a transition between these two alignments which coincides with the resonance position. As a consequence, planetesimals near the resonance point are not well aligned with each other, especially for smaller sizes, thus increasing their velocity dispersion (as we will further discuss in Section \ref{sec:velocity}).

Finally, we also note the occurrence of a second secular resonance near the inner edge of the disk. This is due to the sharp positive slope of the surface density at this edge which contributes a positive component to the precession rate of the planetesimals; see Fig. \ref{fig:fig1}. This additional resonance is not important for our purposes as it transports very little material (see the black line at the inner edge of panels B1 and B2).

The fiducial case was chosen because the timescale for the migration of its planetesimals remains shorter than the timescale for the movement of its resonance (for the first 10 Myrs). As such, the planetesimals can ride the wave of the secular resonance which results in a very efficient transport of material. Despite this, the fiducial case is not the case that transports the most material in our simulations  as some of the bigger planetesimals are left behind the wave ($\sim$ 100 km) (see Fig. \ref{fig:2} A3). To better understand why this happens, next we will explore the condition for transporting planetesimals in more detail.

\subsection{Condition for transport} \label{sec:condition}

\begin{figure}[]\label{fig:3}

\hspace{0.085cm}
\includegraphics[width=8.5cm]{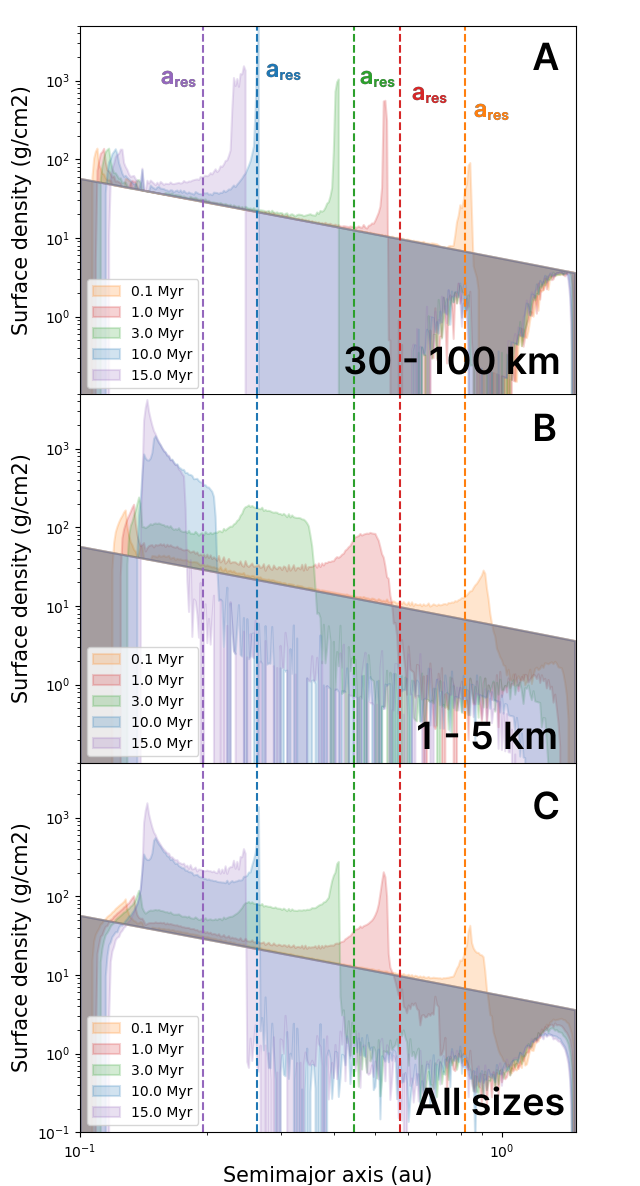}
\caption{Radial profiles of the surface density distribution of planetesimals. In each panel, results are shown relative to the initial profile for different times: 0.1, 0.5, 3, 10, and 15 Myrs.
The vertical dashed lines correspond to the positions of the secular resonance at each considered timestep. \textit{Panel A} shows only the big planetesimals from 30 to 100 km. These planetesimals need to be near the resonance to migrate and when the resonance moves too quickly, the wave cannot catch up on time. \textit{Panel B} shows only the smaller planetesimals from 1 to 5 km. Note that their migration happens more or less irrespective of the resonance position. \textit{Panel C} shows all  considered sizes, i.e., from 1 to 100 km. 	
}
\end{figure}

We can obtain an approximate condition for the effective transport of material. For a planetesimal to remain ahead of the wave, its migration timescale $\tau_{\rm mig}$, at the peak of the eccentricity ($e_{\rm peak}$), must be shorter than the timescale for the movement of the resonance position $\tau_{\rm res}$. As we show in Appendix \ref{app:timescale}, this translates into the following criterion:
\begin{eqnarray} \label{eq:timescale_text}
    &&\tau_{\rm res} \gtrsim \tau_{\rm  mig} \approx \frac{1}{2} \tau_0^{-1/2} \left( A_{\rm pJ}e_{\rm J} \right) ^{-3/2} \nonumber \\
    \approx&& 40 {\rm Myr} \left( \frac{M_c}{1 M_{\rm \sun}} \right) \left( \frac{3 M_{\rm Jup}}{M_{\rm J}} \right)^{3/2} \left( \frac{\Sigma_{\rm gas}}{10 \rm g/cm^2} \right)^{1/2}
    \left( \frac{0.1}{h} \right)^{1/2} \nonumber \\
   &&  \left( \frac{s}{100 \rm km} \right)^{-1/2} \left( \frac{a_{\rm J}}{3 {\rm au} } \right)^{6} \left( \frac{a_{\rm p}}{0.25 {\rm au}} \right)^{-9/2} \left( \frac{e_{\rm J}}{0.05} \right)^{-3/2}.
\end{eqnarray}
where s is the radius of the planetesimal. In order to illustrate the use of this expression,  in Fig. \ref{fig:3} we observe the transport of material and pay attention to the time at which the transport stops working. Panel \ref{fig:3}A shows that the large planetesimals (30 - 100 km) need the resonance to move inwards, while stopping their migration if at any moment the resonance runs past them (i.e. they no longer fulfill Eq. (\ref{eq:timescale_text})), leaving the planetesimals behind as can be seen for the 15 Myrs snapshot. At 10 Myrs, the migration rate of the 100 km planetesimals is around 40 Myrs according to Eq. (\ref{eq:timescale_text}), which is above the timescale for the resonance movement ($\sim \Delta t \times a/\Delta a \sim \rm 5Myrs \times 0.2au/(0.05au) \sim 20 Myrs$).

On the other hand, the small planetesimals (1 - 5 km) all migrate regardless of the position of the resonance because a forced eccentricity below the peak value suffices to drive efficient migration (see Fig. \ref{fig:3}B). Because of our choice of p in Eq. (\ref{eq:planetesimals}), all sizes contribute the same amount of mass to the surface density, and so we still see migration of material even when the resonance moves too quickly (albeit, less efficiently), as can be seen in Fig. \ref{fig:3}C. 

The difference in the migration timescales for different sizes translates into a segregation of planetesimals. This segregation helps to keep the planetesimals better aligned, as when comparing two planetesimals identical in size, their relative velocity would be zero.

\subsection{Amount of transported material} \label{sec:amount}
We now quantify the amount of material transported by the sweeping secular resonance. We do this
by simply integrating the surface density of solid material in places where there is an excess compared to the starting density, i.e.:

\begin{equation} \label{eq:material}
    M_{\rm t} = \int_{\rm r_{\rm in}}^{\rm r_{\rm trans}} \left( \Sigma_{\rm p} - \Sigma_0 \right)  dA = \int_{\rm r_{\rm trans}}^{\rm r_{\rm out}} \left( \Sigma_0 - \Sigma_{\rm p} \right)  dA,
\end{equation}
with $r_{\rm in}$ and $r_{\rm out}$ the limits for the planetesimal distribution, and $r_{\rm trans}$ (the transition point at $\Sigma_{\rm p} = \Sigma_0$) where we go from having an excess of planetesimals to a deficit, compared to the initial distribution (see Fig. \ref{fig:3}). An approximation to this equation can be made, where we assume $\Sigma_{\rm p} = 0$ for the rightmost expression of Eq. (\ref{eq:material}). In other words, we assume all the region depleted of material (swept by the resonance) has no solid material left and set $r_{\rm trans} = a_{\rm res}$. By doing this, we find the following approximation for the fraction of transported mass:

\begin{eqnarray}
 \label{eq:material_approx}
    f \equiv \frac{M_{\rm t}}{M_{0}} &\approx& \int_{\rm \max(0.1,a_{\rm f})}^{\min(1.5,a_{\rm i})} \Sigma_0 dA \bigg/ \int_{\rm 0.1 au}^{\rm 1.5 au} \Sigma_0 dA \nonumber \\
    &\approx& \frac{\min(1.5 \rm au,a_{\rm i}) - \max(0.1 \rm au,a_{\rm f})}{\rm 1.5 au-0.1 au},
\end{eqnarray}
where $a_i$ and $a_f$ represent the initial and final positions of the resonance, respectively, and $M_0$ is the initial solid mass in the disk. In the fiducial case, this fraction is about 0.73 after 10 Myrs.

Next we use Eq. (\ref{eq:material}) to compare the amount of transported material for different simulations with parameters different from the fiducial. 

\subsection{Dependence on different parameters} 

For consistency, we initiate all disks with the same gas/solid surface density and planetesimal distribution so we can report the efficiency of each simulation as simply the fraction of transported mass. To begin with, we note that the amount of transported mass depends on the movement of the resonance location (Equation \ref{eq:material_approx}). In Appendix \ref{app:resonance} we derive a simple equation for the position of the resonance, finding that to a good approximation one can write:
\ba \label{equation_c_text}
\left( \frac{a_{\rm res}}{a_{\rm J}}\right)^{3/2} = \frac{ 0.02 M_{\rm out} \Delta^3 + 3 M_{\rm in} \Delta^{-3/2} }{M_{\rm J}} , 
\ea
where $M_{\rm in}$ and $M_{\rm out}$ are the masses of the inner and outer disks, respectively, and $\Delta$ is the gap width as defined in Eq. \ref{eq:gap_width}. Looking at Equation (\ref{equation_c_text}), we can see the resonance position depends on the Jovian's mass, as well as the inner and outer disk masses. In what follows we will explore how these parameters affect the transported mass. The inner and outer disk mass evolution will be parameterized by the $\alpha$ viscosity parameters. We will also briefly discuss the dependence of the transported mass on the Jovian's position and its orbital eccentricity.

\begin{figure}

\includegraphics[width=9cm]{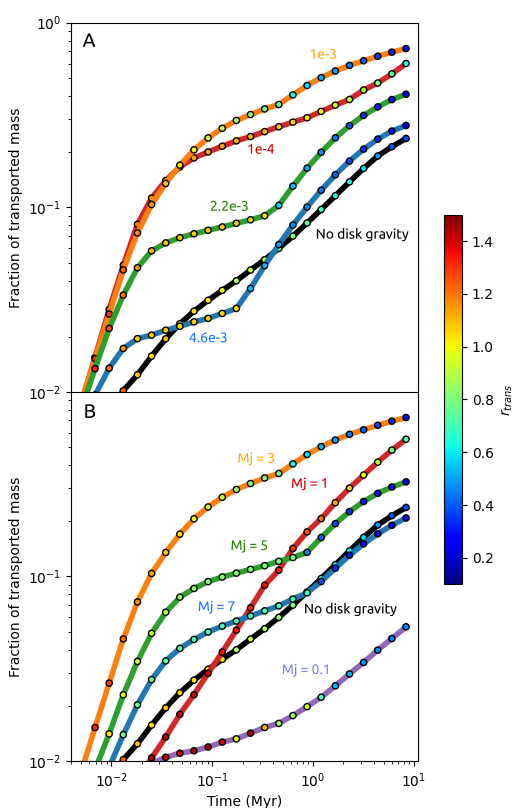}
\caption{Fraction of transported material as a function of time. The total initial mass of planetesimals is 1.8 earth masses, equivalent to 1\% of the gas mass in the inner disk.
Different coloured curves in panel A correspond to different values of $\alpha_{\rm in}$ as highlighted therein; namely, 1$\times 10^{-4}$, 1$\times 10^{-3}$, 2.2$\times 10^{-3}$, and 4.6$\times 10^{-3}$.
Results in panel B correspond to different planetary masses, namely, 0.7, 1, 3, 5, and 7 $M_{\rm Jup}$, and are shown using purple, red, orange, green, and blue curves, respectively. 
In both panels, the black curves represent the case where the gravitational effects of the disk is not taken into account. The colored circle symbols correspond to r$_{\rm trans}$ as defined in Eq. (\ref{eq:material}); see the colorbar. All other planet--disk parameters are set equal to their fiducial values (see Table \ref{table:models}). 
}	
\label{fig:fig4}
\end{figure}

In Fig. \ref{fig:fig4} we vary 2 of these parameters (panel A varying disk viscosities and panel B varying the mass of the Jovian). We will discuss each panel in its corresponding subsections. Also, note that much more material is transported compared to the case in which disk gravity is not accounted for (in Fig. \ref{fig:fig4}, for both panels, compare the orange and black lines, with and without disk gravity respectively, for the fiducial parameters). Next, we will explore the dependence of the fraction of transported mass on each parameter.

\begin{figure}

\includegraphics[width=10cm]{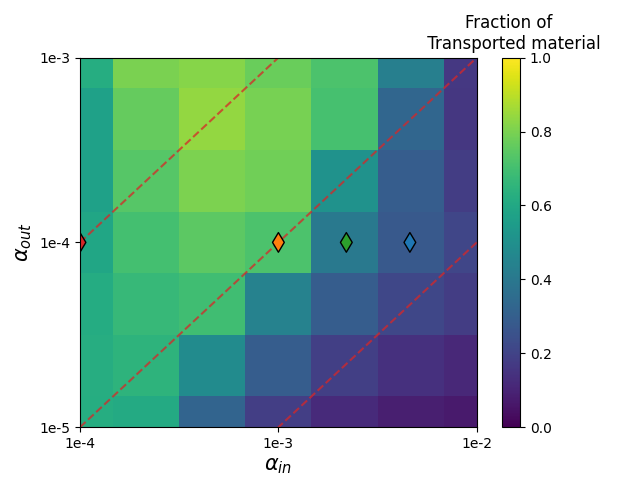}
\caption{Transported mass fraction at the end of the 10 Myr simulation for different combinations of $\alpha_{\rm in}$ and $\alpha_{\rm out}$ 
in simulations which are otherwise identical to the fiducial case (Table \ref{table:models}).
The red lines mark the contours of constant $\alpha_{\rm out}/\alpha_{\rm in}$; with a ratio of 1, 0.1 and 0.01 (top to bottom). The diamond symbols correspond to the same simulations that are depicted in Fig. \ref{fig:fig4}(A) with the same color coding: i.e., with $\alpha_{\rm in}$ = 1$\times 10^{-4}$, 1$\times 10^{-3}$, 2.2$\times 10^{-3}$, and 4.6$\times 10^{-3}$ (shown in red, orange, green, and blue,  respectively).}	
\label{fig:fig5}
\end{figure}

\subsubsection{Dependence on disk viscosities}
\label{sec:dep_viscosity}

In Fig. \ref{fig:fig4}A we can see that higher than fiducial $\alpha_{\rm in}$ values result in less efficient transport as the resonance moves too quickly (see Eq. (\ref{eq:timescale_text})) compared with the fiducial case. On the other hand, for the lower than fiducial $\alpha_{\rm in}$, we would expect the transported mass to catch up if we let the disk evolve for the same number of viscous times as the fiducial. In fact, our results depend on the time we let the disk evolve (10 Myrs in all simulations used for Fig. \ref{fig:fig4}), so the values stated here would actually be a lower limit. If the disk would deplete more slowly (quickly), the resonance would also move slower (faster), thus transporting material more (less) efficiently.

A more exhaustive search of the parameter space for viscosities (following the quoted values for $\alpha$ viscosity parameters listed in Table \ref{table:models}) only showing the final fraction of transported mass can be found on Fig. \ref{fig:fig5} (some simulations from Fig. \ref{fig:fig4}A have been marked on this plot). The peak of transported material in Fig. \ref{fig:fig5} depends on the Jovian mass and in the case of 3 $M_{\rm Jup}$ the simulations transporting the most material are the ones with a ratio $\alpha_{\rm out}/\alpha_{\rm in} \sim $  0.1 - 1. Consistently, these models with $\alpha_{\rm out} < \alpha_{\rm in}$ thought to be more realistic\footnote{Because of the higher irradiation of the star in the inner disk, more ionized material couples with the magnetic field of the star promoting the Magnetorotational Instability and hence higher turbulence which translates into a higher viscosity \citep{gammie,armitage}.}. As we had mentioned previously, if $\alpha_{\rm in}$ is too high, then very little material is transported as the resonance location moves too quickly (which also explains the lower values of $f$ in the right part of the parameter space of Figure \ref{fig:fig5}). In other cases, similar ratios of viscosities give similar fractions of transported material because the inner and outer masses remain proportional.

\subsubsection{Dependence on mass of the giant planet} \label{mass_giant}

We now turn our attention to the dependence of transported mass on planetary mass. 
To this end,  
we consider the evolution of planetesimals in 
otherwise identical disk--planet systems but differing in the value of the planetary mass within the range $[0.1-7] M_{\rm Jup}$ 
; see Table \ref{table:models}. 
The results are summarized in Figure \ref{fig:fig4}(B). 
We generally find that simulations adopting the fiducial planetary mass of 3 $M_{\rm Jup}$ transport the highest fraction of solid material. This can be understood as follows. Increasing the planet's mass results in the resonance condition being established closer to the star, while making the planet smaller makes the resonance location move closer to the planet where there is no solid material in our simulations (see Section \ref{sec:mmr}). Both of these effects reduce the transported mass by reducing the range of solid material swept by the resonance.

\begin{figure}

\includegraphics[width=9cm]{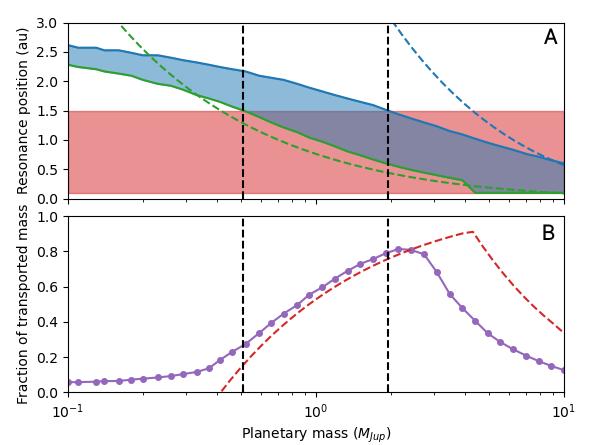}
\caption{The relationship between the radial range over which the resonance sweeps and the fraction of transported mass of planetesimals. 
Panel A shows the range covered by the resonance as a function of $M_{\rm J}$, starting at the blue solid line and ending at the green solid one. 
The approximate initial and final resonance positions as obtained using Eq. (\ref{equation_c_text}) are shown with blue and green dashed curves, respectively.
Panel B shows the amount of transported material, both as measured from the simulations (purple curve) and as 
estimated 
using Eq. (\ref{eq:material_approx}) and (\ref{equation_c_text}) (shown using red dashed curve). See the text (Section \ref{mass_giant}) for details.}	
\label{fig:fig_6}
\end{figure}

Given the strong dependence of the fraction of transported mass on the planet's mass, we provide a more detailed picture in Figure \ref{fig:fig_6}(B).
It is evident that the transported mass peaks at about 2 or 3 Jupiter masses. To explain this we need to look at the range that the resonance sweeps over.

We approximate the position of the resonance from the simulations (blue and green solid curves for initial and final positions, respectively) using Eq. (\ref{equation_c_text}) (blue and green dashed lines) and the transported material with Eq. (\ref{eq:material_approx}) (purple line on panel B). The result is shown as the red dashed line in panel B and provides a rough description of the the transported mass. Three regions can be distinguished in both panels. (i) $M_{\rm J} \lesssim 0.5 M_{\rm Jup}$: the final resonance location is outside the region with material ($0.1 - 1.5$ au) and an insignificant amount of material is transported; (ii) $0.5 M_{\rm Jup} \lesssim M_{\rm J} \lesssim 2 M_{\rm Jup}$: the fraction increases as the resonance sweeping region gets in until it peaks when the starting resonance position enters the region with material, (iii) $M_{\rm J} \gtrsim 2 M_{\rm Jup}$: as the initial resonance location approaches the inner edge of the disk, the material transported goes down again. This simple model we describe in detail in Appendix \ref{app:resonance} together with Eq. (\ref{eq:material_approx}) gives us the same qualitative behavior as full simulations. This includes the upward and downward slope and the peak in the Jovian's mass at around 3 $M_{\rm Jup}$ as can be seen in Fig. \ref{fig:fig_6}B.

\begin{figure}

\includegraphics[width=9cm]{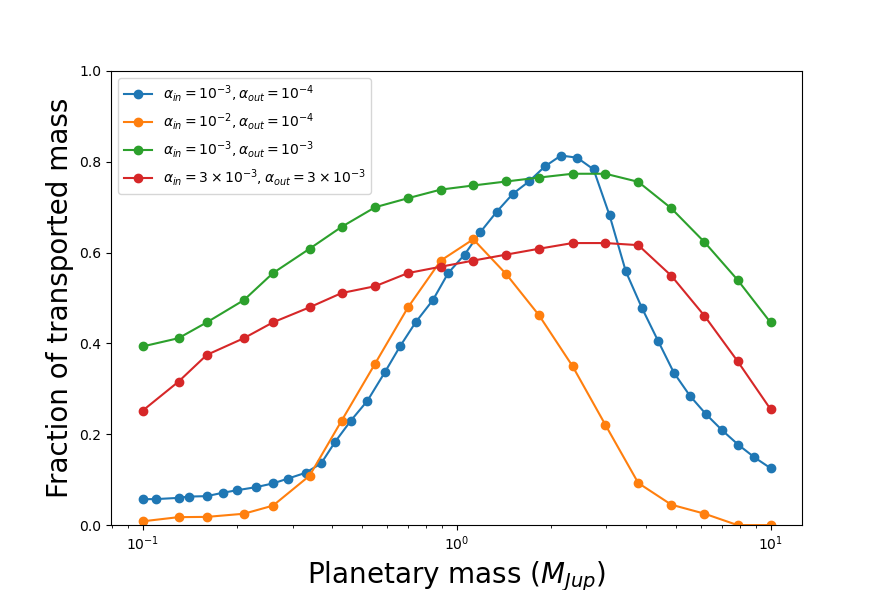}
\caption{Fraction of transported mass of planetesimals at the end of the 10 Myr simulation as a function of Jupiter masses for different combinations of inner and outer viscosities as shown in the legend. The blue curve corresponds to the fiducial viscosities (Table \ref{table:models}).}	
\label{fig:fig_new}
\end{figure}

Furthermore, we can see in Fig. \ref{fig:fig_new} that this shape, including the peak at around $\sim 1-3$ Jupiter masses, is very similar for many combinations of inner and outer viscosities. This is appealing, especially given that the typical measured masses of cold Jupiters in exoplanet systems fall within this range (e.g. \cite{california_2}). 
Finally, to explain the slight difference in shape for the green and red curves shown in Fig. \ref{fig:fig_new} at low Jupiter masses, we note that the final position of the resonance for these cases are already inside the region with solid material (because the outer region gets depleted faster in these cases), so we do not see a first change in slope as with the blue and orange curves.

\subsubsection{Dependence on other parameters} \label{sec:others}

As mentioned before, other parameters can affect the transport of material. The most important ones, besides the parameters mentioned in the previous subsections, are the Jupiter's eccentricity and position.

According to our simple model described in Appendix \ref{app:resonance}, the position of the resonance scales with the Jovian's position (Equation \ref{equation_c_text}). In this sense, we would expect the system to just scale with the position of the Jupiter. However, the available solid material reservoir would be different. Also, the resonance position movement timescale will be affected. Namely, if the resonance starts with the Jovian further away, it will travel faster through the inner disk.

In the case of the eccentricity, from Eq. (\ref{eq:peak_ecc}), one can see that the eccentricity peak of the resonance scales with the Jovian's eccentricity. This will directly impact the migration timescale as can be seen from Eq. (\ref{eq:timescale_text}). A lower Jovian's eccentricity makes the migration timescale longer, hindering the transport. In fact, in the limit of a circular orbit, the Jovian would stop secularly exciting eccentricity onto the planetesimals. We have tested this by lowering $e_J$ to 0.01 and observed little transport, and only of the smallest planetesimals. However, we expect that gap-opening Jovians may retain some eccentricity due to disk-planet interactions with values around 0.07 to 0.09 according to simulations by \cite{Duffell_2015}.

We further discuss the effect of other parameters not mentioned here such as the disk evolution time in Section \ref{sec:discussion}.

\section{Results: Velocity Dispersion}
\label{sec:velocity}

In the previous section, we demonstrated that a significant amount of solid material can be transported from regions close to the giant planet to the inner regions of the disk. In order to asses whether this may result in planet formation we need to calculate the velocity dispersion ($\sigma_v$), which is a function of eccentricity and $\Delta \varpi$, for the planetesimals in each individual ring. This will help us estimate the collisional velocity of this planetesimals once we include collisions in our simulations.

To this end, we first introduce the planetesimal eccentricity vector as follows:
\begin{equation}
    \mathbf{e} = (e \cos \varpi , e \sin \varpi ),
\end{equation}
and an analog to a center of mass velocity
\begin{equation}
    \textbf{e}_{\rm CM} = \frac{\Sigma_{i} \textbf{e}_i m_i}{\Sigma_{i} m_i},
\end{equation}
where i = 1, ..., N for all the planetesimals that fall into a specific ring. Then we compare each individual eccentricity vector of the planetesimals with this center of mass eccentricity using the following relation:
\begin{equation}
    \langle e^2 \rangle = \frac{\Sigma_{\rm i} (\textbf{e}_i - \textbf{e}_{\rm CM})^2 m_i}{\Sigma_{i} m_i}.
\end{equation}
Finally, the dispersion in velocity would be proportional to the dispersion of these eccentricity vectors, such that,
\begin{equation} \label{eq:velocity}
    \sigma_v = v_K \sigma_e = v_K \sqrt{\langle e^2 \rangle}.
\end{equation}

\begin{figure} \label{fig:7}
\includegraphics[width=9cm]{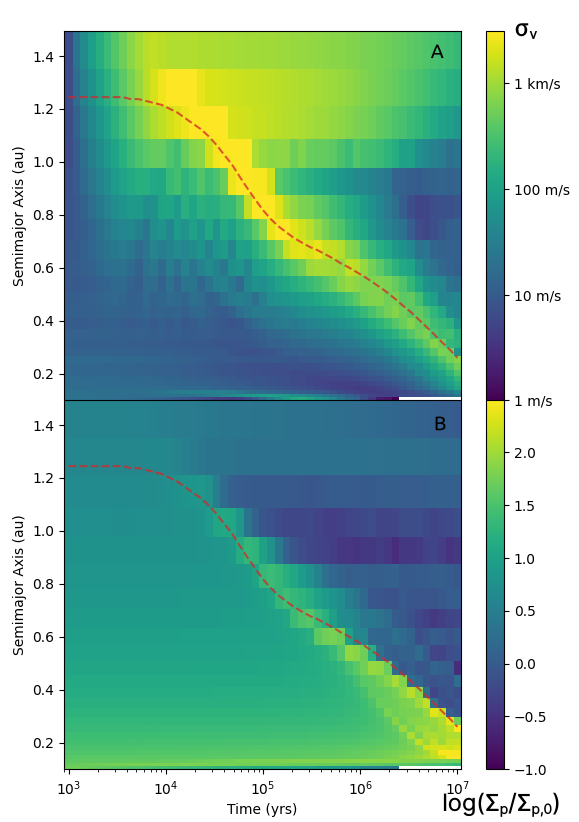}
\caption{Panel A shows the velocity dispersion of planetesimals computed according to Eq. (\ref{eq:velocity}) as a function of time and semimajor axis, using results of the fiducial simulation (Table \ref{table:models}). The red line marks the position of the resonance. Note that the velocity dispersion increases near the resonance due to eccentricities being higher there and also the apsidal angle alignment varies quickly near the peak as can be seen in Fig \ref{fig:2}c. Panel B shows the corresponding surface density of solids compared to the initial solid distribution. Note that most of the material is near the resonance and the surface density interior to the resonance increases with time.}	
\end{figure}

\begin{figure} \label{fig:8}
\includegraphics[width=8cm]{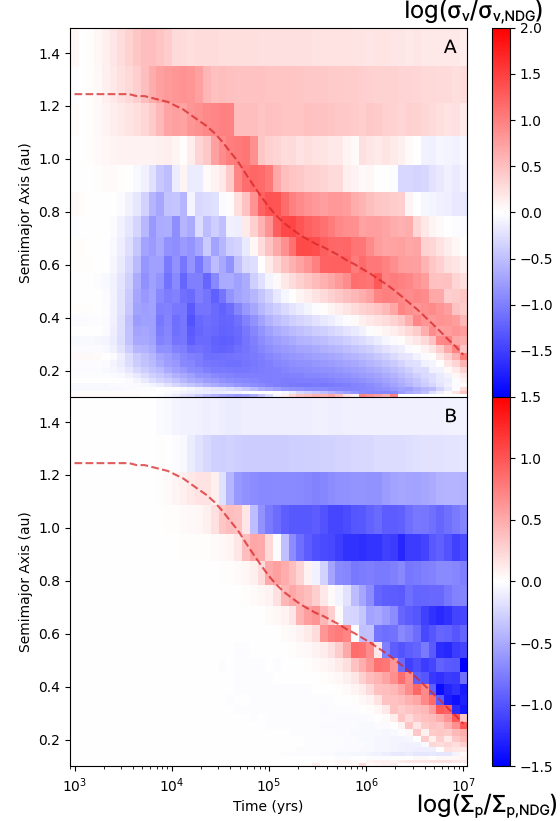}
\caption{Same as Fig. \ref{fig:7} but comparing the fiducial case with the case where no disk gravity (NDG) was considered. Both in log scale.}
\end{figure}

In Fig. \ref{fig:7}A we show the velocity dispersion for our fiducial case as a function of time and semimajor axis. We observe that in regions near the resonance, $\sigma_v$ increases as the eccentricity becomes larger when they approach $e_{\rm peak}$. Additionally, there is a region of misalignment surrounding this position (see Fig. \ref{fig:2}c). Inner to this region, the velocities can get as low as $\sim 10-100$ m/s (which is lower than if the effect of the disk gravity was not included; see e.g. Fig. \ref{fig:8}). This lower velocity dispersion is also due to the segregation in sizes of the planetesimals (as seen in Fig. \ref{fig:3}), because planetesimals of similar size tend to be more aligned. It is worth noting that even if these higher velocity dispersion near the resonance result in destructive collisions, the smaller fragmented planetesimals 
would migrate faster and accumulate in the region with a lower velocity dispersion. We also include the surface density of solids in Fig. \ref{fig:7}B as a function of time and semimajor axis compared to the initial distribution of solids. Here we can see that most of the material concentrates around the resonance, the region swept by the resonance gets depleted and accumulated ahead.

In general, Fig. \ref{fig:7} shows that relative collisional velocities are around 10-100 m/s where material is being accumulated ahead of the resonance. This implies collisions might lead to accretion and not fragmentation in these regions for strong basalt rocks and even weaker rocks (see \citet{collisions}). These values are generally lower than those obtained in the case of a non-self-gravitating disk by about a factor of $\sim 10$ (see Fig. \ref{fig:8}), further demonstrating the role of disk gravity for the transport of solid material to the inner disk.


\section{Discussion} \label{sec:discussion}

We have shown that, by including the effect of disk gravity for an evolving gaseous disk, a large fraction of the solid material can be transported. This transporting effect seems to be very robust, as it works with a wide range of typical Jovian masses and disk viscosities. This effect is also robust against different choices for the planetesimal distribution.

In the following, we first discuss the relevance of including the disk gravity in an evolving disk in Section \ref{sec:relevance}. Then, what implications our mechanism has on inner planet formation in Section \ref{sec:inner_planet_formation}. We mention some caveats in Section \ref{sec:caveats} and finally, some future work we hope to include for the next paper in Section \ref{sec:future_work}.

\subsection{The importance of disk gravity in an evolving disk} \label{sec:relevance}

One crucial ingredient in our model is the inclusion of the disk gravity. To highlight its importance, we now compare our fiducial results with the case where disk gravity is not included which, from hereon, we refer to as the NDG case. In this case there is no secular resonance, and so, only the smaller planetesimals would migrate. In Fig. \ref{fig:8}A we compare the velocity dispersion of planetesimals (according to Eq. (\ref{eq:velocity})) in the fiducial case to that in the NDG case. We can see the peak in eccentricity promotes higher dispersion velocities near the resonance, but away from it, lower  
velocity dispersions of planetesimals
can be achieved
compared to the NDG case. This is appealing, because previous works which did not include disk gravity in their calculations (e.g. \cite{kortenkamp}, \cite{GUO,Guo2023}) do already 
report on the planet-formation-friendly environment that the Jovian, together with the gas drag, could produce. In this context, the inclusion of disk gravity, which further reduces the velocity dispersion, would further enhance the possibility of planet formation.

The effects of this disk-induced secular resonance have also been studied in the shake-up model for the Solar System \citep{Nagasawa_2005}, as well as in the context of planet formation in and around binaries \citep{binaries_1,binaries_2,binaries}.
It is worth noting that these studies concentrate more on the 
coupled eccentricity--apsidal angle evolution of planetesimal orbits,
but not on their transport. Additionally, these works either do not include the evolution of the disk or use a simplistic model (e.g., a uniformly- decaying power-law disk). Instead, our work shows that the evolution of the disk is crucial to predict the evolution of the position of the secular resonance and the subsequent transport of material. There are also works which use N-body simulations to reproduce the cold Jupiter - super Earth correlation (e.g. \citet{bitsch_2023}), but which neglect the effect of disk gravity.

\subsection{Relevance to inner planet formation} \label{sec:inner_planet_formation}
Coming back to what we set off to explore, namely an explanation for the correlation observed between cold giants and super-Earths. We have proven that, thanks to the effect of planetesimal transport by the sweeping of the secular resonance, a significant amount of material with small velocity dispersion is accumulated in the inner region of the disk ( $\lesssim$ 0.3au). Additionally, the transported solid mass is segregated in rings by size. All of this occurs by the time the surface density of gas is largely depleted ($\leq10 {\rm 
 g/cm^2}$, Fig. \ref{fig:2}) which could promote the formation of planets as seen in \citet{morbidelli}. This environment would be ideal for planetesimals to assemble into super-Earths \citep{lee2016}. Another possibility would be to form warm or hot Jupiters, if the transported solid material is sufficient and there is still enough gas to accrete \citep{hot_insitu_1,hot_insitu_2}. This may be the case as the population of hot Jupiters generally have external companions \citep{hotJupiterStatistics}, similar to the close-in Super-Earths. 
 
 It is also possible that the material transported results in a cascade of collisions (especially for planetesimals at and around the resonance where the eccentricites are relatively large). This could have the effect of grinding the material down which then would migrate faster producing a ring of dust or a recycled flux of pebbles that would affect subsequent planet formation by forming planetesimals via streaming instability \citep{drazkowska}.

 We defer a detailed anaylisis of these possibilities to a future work and next discuss the implications of our mechanism for exoplanet systems and the Solar System.

\subsubsection{Implications to exo-planetary systems}

Once we implement collisional evolution of planetesimals, we could start making more specific predictions about the correlation between inner planets and outer giants. For example, the California Legacy Survey \citep{correlation_2021} found that this correlation breaks down if they consider outer giants more massive than about 0.38 M$_{\rm J}$ and closer to the star (0.3–3 au). Similar studies, including the Kepler Giant Planet Survey \citep{weiss2023} will shed light on the various trends (if any) between the gas giant properties and thoose of inner sub-Neptunes that will allow us to compare with our models.

Eventually, we hope to have a more complete panorama with observations from TESS and GAIA ($\sim100$ systems with super-Earths coexisting with outer Jovians with absolute masses in the range $1-5$ au (Espinoza, Zhu \& Petrovich 2023, in prep). Furthermore, with future observations from JWST on atmospheric compositions, we could better determine where the material that forms the inner planets came from (as the planetesimal composition varies with its position relative to the icelines), providing us again with more constraints for our model.

\subsubsection{Implications to the Solar System}

This mechanism is also relevant for the Solar System as it also has cold giant planets. Our results indicate that, during the gas disk depletion, Jupiter would have transported copious amounts of solid material as it falls within the transported mass peak observed in Fig. \ref{fig:fig_new} (albeit, somewhat to the lower end of the peak). However, we have further checked that accounting for the effects of a second giant planet (i.e. Saturn), puts a constraint on the minimum value the precession rate of Jupiter's orbit can have (see Fig. \ref{fig:fig1}), reducing the range of sweeping for the secular resonance, which reduces the transported mass.

We plan on expanding our analysis by including the effects of a second giant planet in a future paper. This kind of simulation has already been performed by \cite{Bromley_2017} to explain the low mass of Mars and the asteroid belt by focusing on the depletion of material that is swept by the secular resonance. This work, however, makes few remarks on the material that is swept to the inner regions. In our work we also see this effect of depletion left behind after the secular resonance has swept. Nonetheless, we care more about the accumulation of material at and interior to the resonance position.

There is an alternative besides the possibility of this transported solid material forming planets. The planetesimals migrating inwards away from Jupiter into the inner region of the disk could have plowed through earlier formed planets in our Solar System, which could be an explanation of why we do not have a multi-planet compact system of super-Earths as many of the other exoplanetary systems do. This plowing of material moved by Jupiter is very similar to the one described in the Grand Tack model \citep{BL2015}, but without fine tuned positions and migration histories of Jupiter and Saturn. Also this effect may still work after considering collisional evolution unlike \cite{BL2015}  (see \citet{deienno_2020}).

\subsection{Caveats} \label{sec:caveats}

We now discuss some of our model assumptions and caveats and how we expect these to impact our results.

\subsubsection{Planetary migration}
\label{sec:planetary_migration}

\begin{figure}[]\label{fig:fig_migration}
\hspace{0.085cm}
\includegraphics[width=10cm]{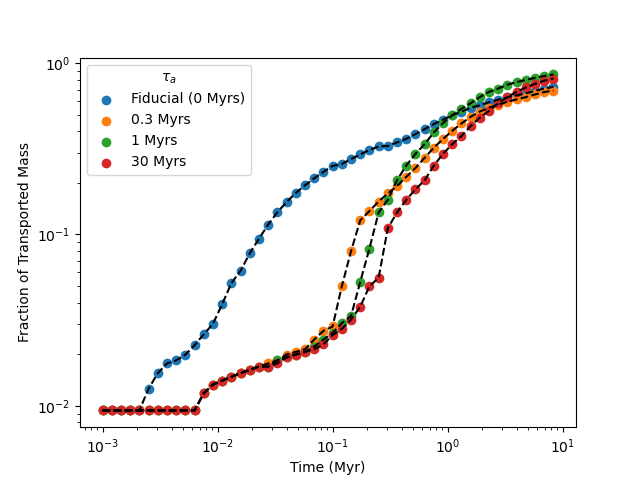}
\caption{Fraction of transported mass in planetesimals as a function of time in systems that are identical to our fiducial model, except in that they assume a migrating giant planet. Different colours correspond to different migration timescales, as indicated in the legend. For further details, see Section \ref{sec:planetary_migration}.}
\end{figure}

Throughout our work, we ignored the potential effects of planetary migration, which, in principle, could be induced by interactions with the gas \citep{migration1,migration2}. Within the framework of our orbit-averaged model, planetary migration  could introduce a time dependence for the position of the secular resonance. In order to test this effect, we repeated the fiducial case simulation by forcing the Jovian to migrate from 5 au to 3 au exponentially with various migration timescales defined as $\tau_a \equiv |\dot{a_{\rm J}}/a_{\rm J}|^{-1}$. The gap produced by the giant planet is also evolved accordingly following the prescription described in Section \ref{sec:gap}. Our results are presented in Fig. \ref{fig:fig_migration}, where we plot the time evolution of the fraction of transported mass for different values of $\tau_a$. Fig. \ref{fig:fig_migration} shows that, when compared with the fiducial simulation with fixed $a_p$ (i.e., $\tau_a = 0$), the fraction of transported mass in planetesimals is very similar, irrespective of the migration timescale: the only difference being a delay in the onset of transported material of about 0.1 Myrs.

We also tested the effect that the planetary migration alone can have considering only the gravitational effects of the planet, i.e., without considering the precession induced by the disk. In this case, the efficiency for transport remains very low, comparable to the fiducial case without disk gravity (as shown in Fig. \ref{fig:fig4}). This further highlights the role of the disk gravity in our proposed mechanism.

The above discussion is based on the secular dynamics of the system. In principle, there can be some material trapped in mean motion resonances (MMRs) -- which, we remind, were ignored in this work (Section \ref{sec:mmr})-- and which would migrate along with the planet. These planetesimals would have their eccentricities resonantly excited as a result of an adiabatic resonance capture (\cite{MD2000}), regardless of the secular resonance. Considering the m:m-1 internal MMRs with the planet (with $m\geq2$), the migration from $a_{J,i}$ and $a_{J,f}$ leads to an eccentricity excitation  of the planetesimal given by \citep{MD2000}: 
\ba \label{eq:mmr}
e_{p}\simeq \sqrt{ \frac{2}{m-1} \left[\left( \frac{a_{J,i}}{a_{J,f}}\right)^{1/2}-1\right]},
\ea
where the planetesimal started from a circular orbit. This would be the maximum reachable eccentricity as it assumes that the capture occurred exactly when the migration started at $a_J=a_{J,i}$, while in reality most planetesimals will be swept at a later stage in the course of the Jovian's migration.
For 2:1 and 3:2 MMRs, Eq. \ref{eq:mmr} implies a maximum eccentricity of $\sim 0.7$ and $\sim 0.5$, respectively. Plugging these values into Equations (\ref{eq:migration}) and (\ref{eq:ecc_evolution}), we then estimate that these planetesimals would spiral in due to drag down to $\sim 1.2$ au and $\sim 1.8$ au for 2:1 and 3:2 MMRs, respectively. Accordingly, this process could bring some new material to the inner region which is not being considered in our simulations as presented in Section \ref{sec:transport}. Therefore, the coupled effects of MMRs and a migrating planet, could enhance the fraction of transported material, but this would still be dominated by the planetesimals that are migrating due to the secular resonance. This is simply because the secular resonance travels further in, close to 0.2 au, transporting more material.

\subsubsection{Jovian's growth}
A more realistic simulation would have the Jovian begin as a core and undergo runaway accretion. The timescale for the formation of the core is a few Myrs \citep{pollack,guilera2020}. During this time, the resonance will remain close to the planet so no sweeping will take place. On the other hand, the runaway accretion is too quick ($\sim 10^5 \rm yrs$) for the system to react, so the planetesimals just align themselves with the Jovian once it stops growing. We have checked that including the Jovian's growth in our simulations is equivalent to starting with a partially depleted disk.

\subsubsection{Outside the giant planet}

In our study, we focused on the inner region of the disk, ignoring the outer region, because of the correlation seen between giant cold planets and inner super-Earths. However, there could also be a traveling secular resonance and an accumulation of material in the outer region that could create a second giant planet external to the first one as explored in \cite{GUO}.
This could work together with the dust traps that are known to be created at the edges of giant planets \citep{Eriksson_2022} to form planetesimals and subsequently a second planet outside.

\subsubsection{Beyond the Lagrange-Laplace approximation: MMRs and higher-order terms} \label{sec:mmr}

We have neglected the effect of higher-order terms in eccentricity ($\mathcal{O}[e^4]$), and all terms which included inclinations. In general, the eccentricity rarely increases above $\simeq 0.3$ and it only does so at early times when the resonance is closest to the planet (e.g., panel A1 of Fig. \ref{fig:2}). Furthermore, this eccentricity is already large enough to promote migration of the larger planetesimals. 
Thus, the Laplace-Lagrange (LL) approximation is often a reasonably good approximation. Adding small inclinations (up to second order) would effectively replicate the effect from the eccentricities as it enters in the gas drag  with a similar dependence (Eq. (\ref{eq:migration}) - (\ref{eq:ecc_evolution})) \citep{adachi76}, while remaining decoupled from the eccentricities in the LL approximation.

We have to keep in mind, that as the eccentricity of the planetesimals increase substantially near the secular resonance, it may be different at this position, which would affect our condition for transport.
As seen in \cite{GUO,Guo2023}, planetesimals in regions close to an MMR do not align well in eccentricity nor apsidal angle. So the other resonances could actually grind down material closer to the Jovian, which would then align better and migrate faster (because of its smaller size), so it could help with the transport of material, or the material could get stuck in an MMR resonance. For the fiducial distance chosen of 3 au, we avoid the major MMRs when considering planetesimals interior to 1.5 au.

\subsubsection{Other disk dispersal effects} \label{sec:dissipative}

We presented a very simple model for the evolution of the gas disk. At the end of our simulations, although the inner disk has been almost completely dispersed, we still have a massive outer disk with  $M_{\rm out}\sim$ 5M$_{\rm J}$ (initially 8M$_{\rm J}$), which would need to be dissipated by other means like photo-evaporation or winds once an inner cavity has been formed. In this context, other dissipative effects which mainly affect the inner or outer disk separately (e.g. external irradiation), might have a significant effect on the evolution of the position of the secular resonance. Indeed, as shown in Appendix \ref{app:resonance} the position and movement of the secular resonance depend significantly on the masses of the inner and outer gas disks. 

\subsection{Future work} \label{sec:future_work}

In this work, we have assumed that planetesimals do not interact with each other (neither gravitationally nor collisionally), even within regions of the disk with high surface densities of solids. If we want to actually form the inner planets, it will be necessary to implement the collisions between planetesimals and follow on the collisional outcome aided by the relative velocity of the encounter, focusing mainly on their evolution in size as they accumulate in the inner disk. This will help to separate the cases where we could get super-Earth formation inner to the Jovian's orbit.

Also, throughout this paper we considered all planetesimals to be in the same plane along with the giant planet and the gas disk. A similar term to Eq. (\ref{eq:migration}) can be added to account for inclinations in their evolution and we would need to generalize the velocity dispersion to include inclinations.

On another note, as was already mentioned, the first Jovian might help with the formation of a second giant planet exterior to its orbit. In this sense, including a second Jovian might be interesting, not least for applicability to the Solar System.

\section{Conclusion}\label{sec:conclusions}

In this work, we have studied the secular orbital dynamics of planetesimals interior to a cold Jupiter (semi-major axes $\gtrsim 3$ au) affected by the gravitational and drag forces from a viscously-evolving gaseous disk. 

We find that the planetesimals are efficiently transported from $\sim 1$ au to the inner regions $\sim 0.1-0.3$ au owing to the sweeping of secular resonances and drag for a wide range of disk and cold Jupiter parameters. As a result, most planetesimals interior to the cold Jupiter end up in a high surface density ring (density boosted by $\sim 1-2$ orders of magnitude) in these regions. In contrast, ignoring the disk gravity yields only a modest fraction of transported mass (limited to the smaller size planetesimals).
Our main findings can be summarized as follows:
\begin{itemize}
    \item The inward motion of the planetesimals is most prominent for those which can remain ahead of the inwardly-moving resonance. This favors smaller planetesimals and sets a limit for the migration given by comparing the timescales from the motion of the resonance and the migration of planetesimals (condition given by Eq. [\ref{eq:timescale_text}]).
    
    \item The transport efficiency peaks at cold Jupiter masses of $1-5 M_{\rm Jup}$ for a wide range of disk viscosities. Below this range, the transport efficiency decreases as the resonance only sweeps near the planet, while very massive Jovians start the sweeping already close to the star so almost no material is transported.
    
    \item The planetesimals move inwards in size-segregated rings with aligned eccentricity vectors, thus lowering their, otherwise catastrophically large, velocity dispersion. Although the velocity dispersion remains large near the peak in eccentricity, their values interior to the resonance are lower compared to the simulations that ignore the disk's gravity.
\end{itemize}

These results point to the possibility of a region of the disk well suited for planet formation. The amount of transported material is so enhanced in some cases that it may even lead to the formation of massive cores before the disk has dissipated, triggering runaway accretion (possible formation path for hot and warm Jupiters). Overall, our work establishes the major role of the disk's gravity in the redistribution of planetesimals. Future work including the collisional evolution of these stirred and size-segregated rings is needed to quantify whether the formation of cores is indeed promoted.

\acknowledgements
M.B. acknowledges additional support from the National Agency for Research and Development (ANID) / Scholarship Program / Doctorado Nacional grant 2021 - 21211921 and  CASSACA grant CCJRF2105.
A.A.S. acknowledges support by the Alexander von Humboldt Foundation through a Humboldt Research Fellowship for postdoctoral researchers. A.A.S. also thanks Jihad Touma and the America University of Beirut, Lebanon, for hospitality during the early stages of this work.
C.P. acknowledges support from ANID Millennium Science Initiative-ICN12\_009, CATA-Basal AFB-170002, ANID BASAL project FB210003, FONDECYT Regular grant 1210425, CASSACA grant CCJRF2105, and ANID+REC Convocatoria Nacional subvencion a la instalacion en la Academia convocatoria 2020 PAI77200076.

\appendix

\section{Disk evolution}\label{app:evolution}

\begin{figure}[t!]
    \centering
    \includegraphics[width=10cm]{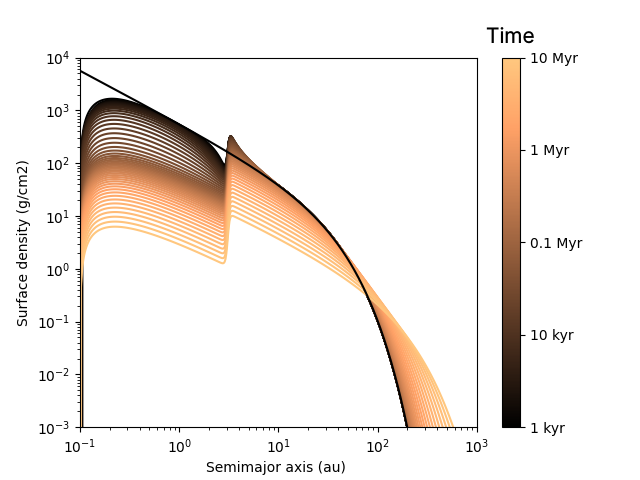}
    \caption{Viscous evolution of the gaseous surface density for 10 Myr for our fiducial case (see Section \ref{sec:fiducial}). Main features are the inner region depleting much faster than the outer disk due to viscosity ($\alpha_{\rm in}=10^{-3}, \alpha_{\rm out}=10^{-4}$). We also see a transient behaviour at the beginning of the simulation located at the boundary of the two regions which depends on the width of the viscosity transition.} 
    \label{fig:A1}
\end{figure}

Throughout our work, following the work of  \cite{guilera2017}, we used a full implicit Crank-Nicholson method to evolve the density profile, solving Eq. (\ref{eq:viscosity}) considering zero torques at the boundaries.
As an example, we show in Figure \ref{fig:A1} the evolution of the surface density for our fiducial two-alpha model (Table \ref{table:models}). We note the development of a peak in the density at the location where the viscosity transition occurs (between $\alpha_{\rm in } = 10^{-3} $ to $\alpha_{\rm out} = 10^{-4}$). This shape can be more pronounced depending on the sharpness of the transition between the viscosities.\footnote{Because no planetesimals go through the transition and the precession rate of the giant planet depends much more on the amount of material outside its orbit (because at its position, gas is severely depleted), this transition is not relevant to our simulations.}
We have also checked that, regardless of the initial slope in the inner disk, the slope of the distribution tends to converge to the same value of about $\gamma = 0.5$ within the first 1 Myr. This observation will become important for our calculations in Appendix \ref{app:resonance} where we estimate the precession rate of the planetesimals.

\section{Migration timescale}\label{app:timescale}

To derive a lower limit for the migration timescale of planetesimals, we need to consider planetesimals at the peak of the forced eccentricity, right at the secular resonance. To begin with, we note that planetesimal orbits exactly at the resonance are perpendicularly aligned with the giant (Fig. \ref{fig:2}). We thus set $\cos \Delta\varpi = 0$ in Equation (\ref{eq:ecc_evolution_3}) which accounts for perturbations due to the gas drag and the Jovian, finding that the peak eccentricity $e_{\rm peak}$ is given by:

\begin{equation}
    \frac{de_{\rm p}}{dt} = A_{\rm pJ}e_{\rm J} - \frac{e_{\rm p}^2}{\tau_0} \sqrt{\frac{5}{8} + \frac{\eta^2}{e_{\rm p}^2}} = 0
\end{equation}

As $\eta \sim \frac{1}{2}h^2 \sim 10^{-4}$ it is safe to assume that $e_{\rm p}^2$ is bigger than $\eta^2$, thus:

\begin{equation}\label{eq:peak_ecc}
    e_{\rm peak} = \sqrt{A_{\rm pJ}e_{\rm J}\tau_0 \sqrt{\frac{8}{5}} }.
\end{equation}
Replacing Eq. (\ref{eq:peak_ecc}) into Eq. (\ref{eq:migration}) gives us the migration timescale:
\begin{equation}
    \frac{1}{\tau_{\rm mig}} = \left \lvert \frac{1}{a_{\rm p}} \frac{da_{\rm p}}{dt} \right \rvert = \frac{2}{\tau_0} \sqrt{\frac{5}{8}e_{\rm peak}^2 + \eta^2} \left[\eta + \left(\frac{\Gamma}{4} + \frac{5}{16}\right)e_{\rm peak}^2\right] \approx \frac{2}{\tau_0} \sqrt{\frac{5}{8}e_{\rm peak}^2} \left(\eta + \frac{1}{2}e_{\rm peak}^2\right).
\end{equation}
Again, assuming, $e_{\rm p}^2$ is bigger than $\eta^2$ we get
\begin{equation}
    \frac{1}{\tau_{\rm mig}} = \sqrt{ \frac{A_{\rm pJ}e_{\rm J}}{\tau_0}} (A_{\rm pJ}e_{\rm J}\tau_0 + h^2) .
\end{equation}
Assuming a planetesimal density of 1 g/cm$^3$ and $h^2 \sim e_{\rm peak}$, we get the following equation considering the parameters of the fiducial case:

\begin{eqnarray} \label{eq:timescale}
    &&\tau_{\rm res} > \tau_{\rm mig} \approx \frac{1}{2} \tau_0^{-1/2} \left( A_{\rm pJ}e_{\rm J} \right) ^{-3/2}  , \nonumber \\
    \approx&& 40 {\rm Myr} \left( \frac{M_c}{1 M_{\sun}} \right) \left( \frac{M_{\rm J}}{3 M_{\rm Jup}} \right)^{-3/2} \left( \frac{\Sigma_g}{10 \rm g/cm^2} \right)^{1/2} 
   \left( \frac{h}{0.1} \right)^{-1/2} \left( \frac{s}{100 \rm km} \right)^{-1/2} \left( \frac{a_{\rm J}}{3 \rm au} \right)^{6} \left( \frac{a_{\rm p}}{0.25 \rm au} \right)^{-9/2} \left( \frac{e_{\rm J}}{0.05} \right)^{-3/2}.
\end{eqnarray}
where, we remind, $s$ is the radius of the planetesimal.

\section{Resonance location: approximate treatment} \label{app:resonance}

To understand how the resonance position $a_{\rm res}$ depends on the parameters of our simulations, we will approximate the two-$\alpha$ disk model by two profiles with total masses of $M_{\rm in}$ and $M_{\rm out}$, respectively and slope of $\gamma$ for both disks. We also denote by $r_{\rm in}$ and $r_{\rm out}$ the inner and outer edges of the gap carved by the planet, respectively. We assume that 
$r_{\rm cut}$, the cutoff radius for the disk, is sufficiently far away from the gap ($r_{\rm cut}/r_{\rm out} \gg 1$).

Let us first consider the giant planet. Since it opens a gap around its orbit, its apsidal precession rate $\dot{\varpi}_J$ can be computed by accounting for the gravity of the inner and outer disks.  Since the latter is more massive, i..e, $M_{\rm out} \gg M_{\rm in}$, it is expected to dominate $\dot{\varpi}_J$. Accordingly, we can employ the disturbing function due to the outer disk, 
with the latter being the dominant  (simply because ). Accordingly, we approximate $\dot{\varpi}_J$ using the disturbing function:
\ba
\mathcal{R_{\rm out}} &=& \left[\int_{\rm r_{\rm out}}^{\rm r_{\rm cut}} \frac{1}{4} a_{\rm J}^2 n_{\rm J}^2 \frac{(2-\gamma) M_{\rm out} r^{1-\gamma}}{r_{\rm out}^2 M_c} \left( \frac{a_{\rm J}}{r} \right)^2 b^{(1)}_{\rm 3/2} (a_{\rm J}/r) du\right] \left[ \frac{1}{2} e_{\rm J}^2 + \mbox{cst.} \right], 
\ea
to arrive at \citep[see e.g.][]{petro2019, srw21}:
\ba
\dot{\varpi}_{\rm J} &=& \frac{3 n_{\rm J}}{4} \frac{M_{\rm out}}{M_c} \left( \frac{a_{\rm J}}{r_{\rm out}} \right)^3 \mathcal{B}(\gamma, r_{\rm out},r_{\rm cut}).  
\ea
Here, $\mathcal{B}$ is a correction term that accounts for approximating the Laplace coefficient $b_{3/2}^{(1)}(\alpha)$ as $\approx 3 \alpha$; it is a weak function of the ratio $r_{\rm cut}/r_{\rm out}$ and the surface density slope $\gamma$. For our fiducial parameters, we find that $\mathcal{B} \approx 0.02$.

Let us now consider a planetesimal embedded in the inner disk. Its orbit will undergo apsidal precession at a rate $\dot{\varpi}_{\rm p}$ determined by the inner and outer disks, as well as the planet. We find that by and large, the contribution the planet dominates over that of the outer disk, and so we neglect the effects of the latter in the following. Accordingly, we can write 
\ba
\dot{\varpi}_{\rm p} &=& A_{\rm pp} + A_{\rm pJ}\frac{e_{\rm J}}{e_{\rm p}}\cos(\varpi_{\rm p} - \varpi_{\rm J}) - \frac{M_{\rm in}}{M_c} \bigg(1-\frac{\gamma}{2}\bigg) \left( \frac{r_{\rm in}}{a_{\rm p}} \right)^{\gamma-2}  n_{\rm p} \mathcal{F} (\gamma,r_{\rm in})  , 
\ea
which can be further simplified by noting that, at the resonance, the cosine of the apsidal angles is zero (Row C of Fig. \ref{fig:2}). Here, the factor $\mathcal{F}$ is proportional to  $\xi (1+\eta_2)$, where $\xi$ is a normalization factor and $\eta_2$ is a correction for a finite disk; see appendix B of \cite{Tamayo_2015}.  For our fiducial parameters, we have $\mathcal{F} \approx 3$.
Finally, taking the limit of $a_p/a_{\rm J} \ll 1$, we arrive at: 
\ba
\dot{\varpi}_{\rm p} &=& \frac{3n_{\rm p}}{4} \frac{M_{\rm J}}{M_c} \left( \frac{a_{\rm p}}{a_{\rm J}} \right)^3 - \frac{M_{\rm in}}{M_c} \bigg(1-\frac{\gamma}{2}\bigg) \left( \frac{r_{\rm in}}{a_{\rm p}} \right)^{\gamma-2}  n_{\rm p} \mathcal{F} 
\ea

A secular resonance occurs at the radial location where the precession rates of the planet and the planetesimals match, i.e., $\dot{\varpi}_p = \dot{\varpi}_J$, which yields the following relation:
\ba
\frac{3}{4} \frac{M_{\rm out}}{M_c} \left( \frac{a_{\rm J}}{r_{\rm out}} \right)^3 n_{\rm J} B = 
\frac{3}{4} \frac{M_{\rm J}}{M_c} \left( \frac{a_{\rm res}}{a_{\rm J}} \right)^3 n_{\rm p} - \frac{M_{\rm in}}{M_c} \bigg(1-\frac{\gamma}{2}\bigg) \left( \frac{r_{\rm in}}{a_{\rm res}} \right)^{\gamma-2}  n_{\rm p} F
\label{eq:res_cond_appAA}
\ea
Assuming that the gap around the planet is of similar width on either side of the orbit such that  $\Delta = a_{\rm J}/r_{\rm out} \approx r_{\rm in}/a_{\rm J}$, and noting that $n_{\rm p}/n_{\rm J} = ( a_{\rm res}/a_{\rm J})^{-3/2}$, Equation (\ref{eq:res_cond_appAA}) can be simplified to read:
\ba
M_{\rm out} \Delta^3 \mathcal{B} =   \left[ M_{\rm J} \left( \frac{a_{\rm res}}{a_{\rm J}}\right)^3 - (4-2\gamma) \Delta^{\gamma-2} \left( \frac{a_{\rm res}}{a_{\rm J}}\right)^{2- \gamma} M_{\rm in} \right] \left( \frac{a_{\rm res}}{a_{\rm J}}\right)^{-3/2},
\label{eq:res_cond_to_solve_appA}
\ea
In Fig. \ref{fig:A1}, we can see that the slope of the surface density becomes shallower as time goes on. When the slope approaches $\gamma=0.5$ at late times, Equation (\ref{eq:res_cond_to_solve_appA}) is easy to solve, which yields:

\ba
\left( \frac{a_{\rm res}}{a_{\rm J}}\right)^{3/2} = \frac{ 0.02 M_{\rm out} \Delta^3 + 3 M_{\rm in} \Delta^{-3/2} }{M_{\rm J}}. 
\ea
We note that a better approximation of the position of the resonance can be obtained if we account for the exact surface density of the inner disk (rather than approximating it as a simple power-law). Indeed, at around $r \approx 1-1.5$ au, the index $\gamma$ increases due to the giant planet's gap, and in the range $\approx 0.1-0.2$ au, $\gamma$ becomes negative as the slope inverts (see Fig. \ref{fig:2}). Additionally, we can use the approximation for the gap width in Eq. (\ref{gap}) from \cite{Kanagawa2016} to define a more accurate value for $\Delta$, such that:

\ba \label{eq:gap_width}
    \Delta = 1 - \frac{\Delta_{\rm gap}}{2 a_{\rm J}} \approx 0.85.
\ea
This completes our derivation of an approximate expression for $a_{\rm res}$.

\bibliography{refs}

\end{document}